\documentclass[aps,prb,twocolumn,superscriptaddress, showpacs]{revtex4}

\usepackage{amsmath,amssymb}
\usepackage{graphicx}
\usepackage{bm}
\usepackage{epstopdf}
\usepackage{color}

\newcommand{\slrr}      {$T_1^{-1}$}

\newcommand{\cu}        {$^{63}$Cu}
\newcommand{\cuf}        {$^{65}$Cu}

\newcommand{\srcuo}{SrCuO$\mathrm{_{2}}$}
\newcommand{\srcacuo}{Sr$\mathrm{_{0.9}}$Ca$\mathrm{_{0.1}}$CuO$\mathrm{_{2}}$}
\newcommand{\srcuod}{Sr$\mathrm{_{2}}$CuO$\mathrm{_{3}}$}
\newcommand{\srcacuod}{Sr$\mathrm{_{1.9}}$Ca$\mathrm{_{0.1}}$CuO$\mathrm{_{3}}$}

\begin{document}

\title{Spin Gap in the Single Spin-1/2 Chain Cuprate \srcacuod}

\author{F. Hammerath}
\email[Corresponding author: ]{f.hammerath@ifw-dresden.de}
\affiliation{IFW-Dresden, Institute for Solid State Research, PF
270116, 01171 Dresden, Germany}
\affiliation{Dipartimento di Fisica and Unit\`a CNISM di Pavia, I-27100 Pavia, Italy}
\author{E. M. Br\"uning}
\altaffiliation[Present address: ]
{Berlin Institute of Technology, Department of Energy Engineering, Energy Process Engineering and Conversion Technologies for Renewable Energies, Berlin, Germany.}
\affiliation{IFW-Dresden, Institute for Solid State Research, PF
270116, 01171 Dresden, Germany}
\author{S. Sanna}
\affiliation{Dipartimento di Fisica and Unit\`a CNISM di Pavia, I-27100 Pavia, Italy}
\author{Y. Utz}
\affiliation{IFW-Dresden, Institute for Solid State Research,
PF 270116, 01171 Dresden, Germany}
\author{N.~S.~Beesetty}
\affiliation{SP2M-ICMMO UMR-CNRS 8182, Universit\'{e} Paris-Sud, 91405 Orsay CEDEX, France}
\author{R. Saint-Martin}
\affiliation{SP2M-ICMMO UMR-CNRS 8182, Universit\'{e} Paris-Sud, 91405 Orsay CEDEX, France}
\author{A. Revcolevschi}
\affiliation{SP2M-ICMMO UMR-CNRS 8182, Universit\'{e} Paris-Sud, 91405 Orsay CEDEX, France}
\author{C. Hess}
\affiliation{IFW-Dresden, Institute for Solid State Research, PF
270116, 01171 Dresden, Germany}
\author{B. B\"{u}chner}
\affiliation{IFW-Dresden, Institute for Solid State Research,
PF 270116, 01171 Dresden, Germany}
\affiliation{Institute for Solid State Physics, Dresden Technical University, TU-Dresden, 01062 Dresden, Germany}
\author{H.-J. Grafe}
\affiliation{IFW-Dresden, Institute for Solid State Research, PF
270116, 01171 Dresden, Germany}
\date{\today}


\begin{abstract}

We report $^{63}$Cu nuclear magnetic resonance and muon spin
rotation measurements on the S=1/2 antiferromagnetic Heisenberg
spin chain compound \srcacuod. An exponentially decreasing
spin-lattice relaxation rate \slrr\ indicates the opening of a
spin gap. This behavior is very similar to what has been observed
for the cognate zigzag spin chain compound \srcacuo, and confirms
that the occurrence of a spin gap upon Ca doping  
is independent of the interchain exchange coupling
$J'$. Our results therefore generally prove the appearance of
a spin gap in an antiferromagnetic Heisenberg spin chain induced
by a local bond disorder of the intrachain exchange coupling $J$. A low
temperature upturn of \slrr\ evidences growing magnetic
correlations. However, zero field muon spin rotation measurements
down to 1.5\,K confirm the absence of magnetic order in this
compound which is most likely suppressed by the opening of the
spin gap.

\end{abstract}

\maketitle

\section{Introduction}

Quantum effects in low dimensional magnets lead to a diversity of novel ground states, thereby attracting much experimental and theoretical attention in order to understand the impact of these effects on the properties of low dimensional systems.
Among others, the integrable one-dimensional (1D) antiferromagnetic S=1/2 Heisenberg spin chain is a particularly interesting
model system. It is characterized by a quantum critical ground state lacking any long range order and by a gapless spectrum of elementary S=1/2 excitations (spinons). Although being known since more than 80 years, it is still a very interesting model system from both the theoretical and the experimental point of view. This was demonstrated by a recent report on highly precise measurements of the spinon excitation spectrum of a 1D S=1/2 antiferromagnetic Heisenberg chain material by means of inelastic neutron scattering.\cite{Mourigal2013} The chain cuprates \srcuod\ and \srcuo\ are very good realizations of the 1D antiferromagnetic S=1/2 Heisenberg model.
They exhibit static magnetism only at low temperatures, namely at $T_N = 5.4$\,K and $T_N = 2$\,K, respectively.\cite{Keren1993, Kojima1997, Matsuda1997} For \srcuo, inelastic neutron scattering (INS) measurements reported the short-ranged nature of the static magnetism below $T_N$ and the gapless spinon excitation spectrum down to the lowest measured energy transfer of 0.5\,meV.\cite{Zaliznyak1999}
The $^{63}$Cu nuclear magnetic resonance (NMR) spin-lattice relaxation rates \slrr\ of both compounds display the theoretically predicted temperature independent behavior of a S=1/2 antiferromagnetic Heisenberg chain.\cite{Sachdev1994, Sandvik1995, Takigawa1996, HammerathPRL2011}
Spin-charge separation, which is an essential feature of the 1D Hubbard model,\cite{Giamarchi2004} was experimentally observed already 15 years ago for both compounds.\cite{Kim1996, Neudert1998}
Recently, a spin-orbital separation has been reported for \srcuod\ by means of resonant inelastic x-ray spectroscopy (RIXS).\cite{SchlappaNature2012}

Apart from the experimental evidences for the good one dimensionality of these compounds, up to now only a few investigations focused on the effects of disorder introduced by impurities in these systems. However, such studies can reveal important information. For instance, by increasing the purity of the primary chemicals for the crystal growth, a very large spinon heat conductivity could be observed in highly-pure \srcuo, indicating ballistic transport, which is a consequence of the integrability of the Heisenberg model.\cite{Hlubek2010} Direct magnetic (Ni) and nonmagnetic (Pd) impurity doping on the Cu sites within the spin chain structure of \srcuod\ led to an expected increase of the Curie contribution in the magnetic susceptibility and to the theoretically predicted reduction of $T_N$.\cite{Mahajan2001, Kojima2004} For \srcuo\ it has been shown that the increase of the Curie contribution is the same for magnetic (Ni) and nonmagnetic (Zn, Ga) dopants, suggesting that its origin is the general chain break effect of the impurities, while Ni remains in a low spin (S=0) state.\cite{Chattopadhyay2011} However, the observation of only a slight increase of the Curie contribution upon Ni doping could also be linked to the appearance of a pseudo spin gap in the spinon excitation spectrum of SrCu$_{0.99}$Ni$_{0.01}$O$_2$, as recently reported by inelastic neutron scattering.\cite{Simutis2013}
Unexpected strong effects have also been observed in \srcacuo. 
The introduction of nonmagnetic Ca dopants on the Sr sites outside the zigzag chain structures drastically suppressed the spinon heat transport\cite{Hlubek2011} and led to an exponential decrease of the $^{63}$Cu NMR spin-lattice relaxation rate \slrr\ below 100\,K, pointing towards the opening of a spin gap.\cite{HammerathPRL2011}
Density-matrix renormalization group (DMRG) calculations suggested that Ca induces a weak bond disorder in the spin chains leading to a small
alternation $\delta$ of the intrachain exchange coupling $J$ along the chains which causes the opening of the spin gap $\Delta$.\cite{HammerathPRL2011}
These calculations also showed that the value of $\Delta$ does not depend on the strength and the sign of the interchain coupling $J'$, but on the strength of the alternation $\delta$ of the intrachain coupling $J$.

Here, we report measurements of the $^{63}$Cu NMR spin-lattice relaxation rate \slrr\ on the single chain compound \srcacuod, where we observe a very similar decrease of the $^{63}$Cu NMR \slrr, confirming that the interchain coupling constant $J'$ is indeed negligible. Additionally, we performed zero-field (ZF) muon spin rotation ($\mu$SR) measurements to search for static magnetism. Down to 1.5\,K we did not observe any evidence for the onset of a magnetic order in \srcacuod.


\section{Sample details, sample preparation and experimental details}
\label{sampleprep}

\begin{figure}
\centering
\includegraphics[width=0.8\linewidth]{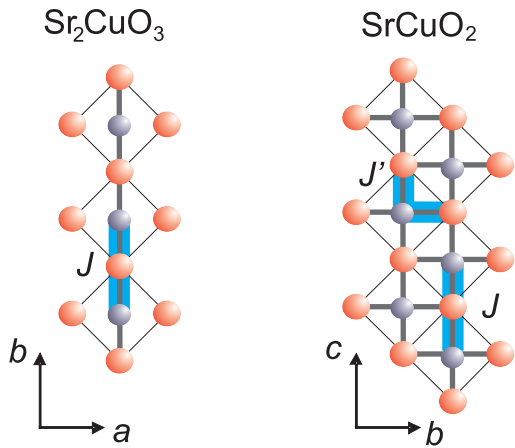}
\caption{\label{fig:chains} Chain structure of \srcuod\ and double
chain structure of \srcuo\ including the main exchange couplings
$J$ (intrachain) and $J'$ (interchain). Grey spheres represent the copper ions, red spheres the oxygen ions. }
\end{figure}

In contrast to the zigzag spin chains of \srcuo,
the crystal structure of \srcuod\ contains only single chains of Cu$^{2+}$ (3$d^9$, S=1/2) ions, running along the crystallographic $b$ direction (see Fig.~\ref{fig:chains}).
The nearest neighbor (NN) intrachain exchange interaction between adjacent Cu moments is antiferromagnetic and its coupling constant amounts to $J \approx (2200 \pm 200)$\,K, which is of the same order as in the zigzag chain compound \srcuo.\cite{Motoyama1996} The only difference between the two compounds is the existence of a weak, ferromagnetic interchain coupling with $|J'/J| \approx 0.1 -0.2$ in \srcuo\cite{Rice1993}, which, due to the linear arrangement of Cu ions, is absent in \srcuod.

Single crystals of \srcacuod\ were prepared using the
traveling-solvent floating zone technique.\cite{Revcolevschi1999}
The crystals cleave readily along (h,0,0) and were verified to
be free from secondary phases by x-ray diffraction measurements.
The stoichiometry was confirmed using EDX and the crystals were
oriented using Laue x-ray back scattering method. Several
previous refinement studies confirm that Ca indeed occupies the Sr
sites in the (Sr$_{1-x}$Ca$_x$)$_2$CuO$_3$ series.\cite{Lines1991,
Xu1992}

\cu\ NMR spectra have been measured with the standard Hahn spin echo method at a constant frequency of 75.742\,MHz by sweeping the magnetic field parallel to the crystallographic $b$ axis. Since \cu\ has a nuclear spin $I=3/2$, its NMR spectrum consists of three quadrupolar split lines, where the central line is affected by first order magnetic hyperfine and second order quadrupolar interactions, while the two satellites are subjects to first order magnetic hyperfine and first order quadrupolar effects.\cite{Takigawa1998}
\cu\ NMR ($I=3/2$) spin-lattice relaxation rate measurements were carried out on the central line of the quadrupolar split \cu\ NMR spectrum (corresponding to the transition $I_z = -1/2 \leftrightarrow I_z = +1/2$) in a slightly higher magnetic field of $\mu_0 H = 7.0493$\,T, with $H$ being parallel to the crystallographic $a$ and $b$ axes. The inversion recovery method was used to measure \slrr. The relaxation function for a magnetic relaxation of a $I=3/2$ nuclear spin system measured at the central transition is given by:\cite{McDowell1995}
\begin{equation}
M_z(t)=M_0\left[1-f \left(0.9 e^{-(6t/T_1)^{\lambda}}+0.1 e^{-(t/T_1)^{\lambda}}\right)\right] \, ,
\label{eq:fit}
\end{equation}
where $M_0$ is the saturation value of the nuclear magnetization, the parameter $f$ is ideally 2 for a complete inversion and $\lambda <1$ accounts for a distribution of spin-lattice relaxation times around the characteristic value \slrr.
At high temperatures, $\lambda \approx 1$ confirms a well defined spin-lattice relaxation rate \slrr.
At lower temperatures, $\lambda$ starts to decrease, indicating a distribution of $T_1^{-1}$ (see inset of Fig.~\ref{fig:T1}).

Zero field (ZF) $\mu$SR measurements down to 1.5\,K have been performed at the GPS instrument of the $\pi$M3 beamline at Paul Scherrer Institute (PSI) in Villigen (CH). To reduce the background, two single crystals of dimensions 5\,x\,4\,x\,1\,mm$^3$ each were used, aligned with their crystollagraphic $c$ axis parallel to the incident muon beam.


\section{Experimental Results and Discussion}

\begin{figure}[t]
\centering
\includegraphics[width=\linewidth]{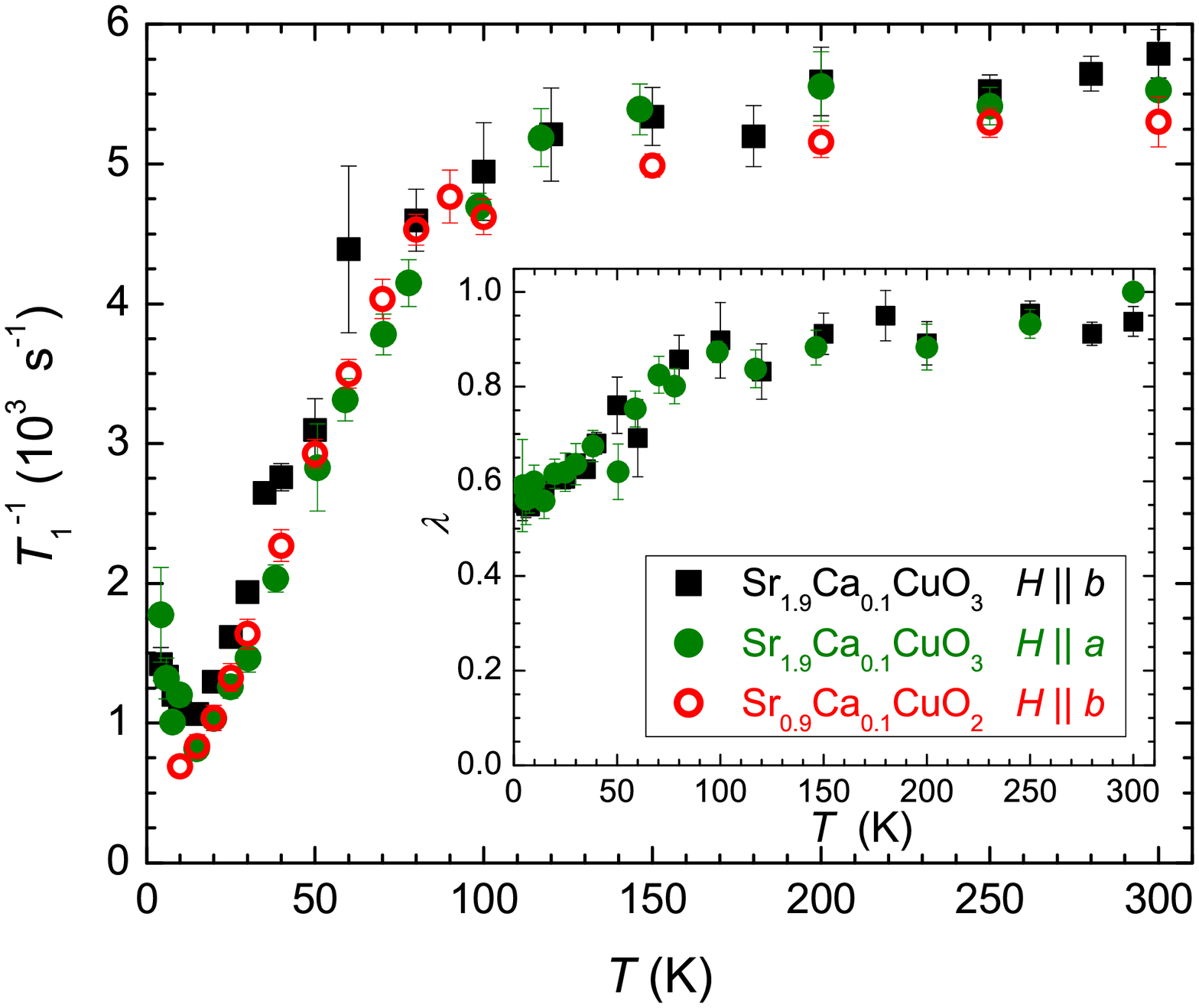}
\caption{\label{fig:T1} Temperature dependent \slrr\ for \srcacuod\ measured along the $b$ (black squares) and $a$ (green dots) direction,
in comparison to \slrr\ of \srcacuo\ measured along $b$ (open red dots, taken from Ref.~\onlinecite{HammerathPRL2011}). Note that the $b$ direction of \srcuo\ is cristallographically equivalent to $a$ of \srcuod\ (see Fig.~\ref{fig:chains}). The inset shows the temperature evolution of the stretching exponent $\lambda$ of the relaxation function (see Eq.~\ref{eq:fit}).}
\end{figure}

Fig.~\ref{fig:T1} shows the temperature dependence of \slrr\ for \srcacuod\ in comparison to the data of the zigzag chain compound \srcacuo.\cite{HammerathPRL2011}
At high temperatures, both compounds display an almost constant \slrr, as it is theoretically expected for 1D antiferromagnetic S=1/2 Heisenberg chains\cite{Sachdev1994, Sandvik1995} and has been reported for the undoped parent compounds.\cite{Takigawa1996, HammerathPRL2011}
Below $T\approx 90$ K, \slrr\ of \srcacuod\ decreases exponentially. This decrease is identical to the one observed in the Ca doped zigzag chain compound \srcacuo\ and points towards the opening of a pseudo spin gap of the same order of magnitude ($\Delta = 50$\,K) as in \srcacuo.\cite{HammerathPRL2011}

On the one hand, this result indicates that the double chain structure of \srcuo\ and the associated interchain coupling constant $J'$ are not necessary prerequisites for the appearance of a gap in the spin excitation spectrum of the spin chains. This has also been reported by our previous DMRG calculations, which showed  that the value of the spin gap does not depend neither on the value nor on the sign of the interchain coupling $J'$.\cite{HammerathPRL2011} On the other hand, these DMRG calculations also showed, that the spin gap is a consequence of a subtle bond disorder induced by the Ca doping, and that the value of the spin gap $\Delta$ depends sensitive on the value of the alternation $\delta$ of the intrachain coupling $J$. In this regard, the perfect agreement between the two data sets pointing towards spin gaps of equal sizes is surprising, since $\delta$ should depend delicately on the structural background (e.g. elastic constants) which is supposed to differ between the two systems. The observation of the very same decrease of \slrr\ upon cooling might thus either be a coincidence or point towards an intrinsic energy scale defined by a collective ordering phenomena in disordered spin chains. Therefore, further studies (e.g. of different doping levels) are envisaged.

\begin{figure}[t]
\centering
\includegraphics[width=\linewidth]{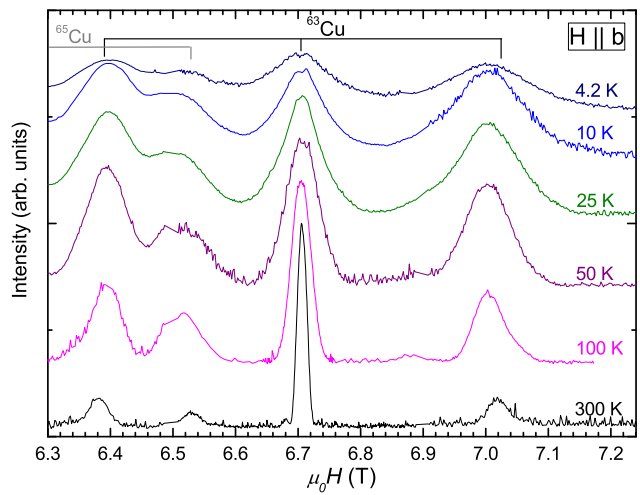}
\caption{\label{fig:spec} \cu\ NMR spectra of \srcacuod\ for
$H \parallel b$ measured at different temperatures (see labels). All three resonance lines of the quadrupolar split \cu\ 
NMR spectrum are visible. Additionally, the high field satellite of the \cuf\ isotope is visible at $\sim$ 6.52\,T.}
\end{figure}

Apart from the very similar decrease of \slrr, there are also slight differences between the two data sets of \srcacuo\ and \srcacuod. At first, \slrr\ of \srcacuod\
starts to increase again below $T\approx$\,15\,K (see Fig.~\ref{fig:T1}). This upturn could indicate a
slowing down of magnetic fluctuations at least in some parts of the
chains. Secondly, below $T\sim$ 80\,K the relaxation rates become
more distributed, indicated by a decrease of the stretching exponent $\lambda$ (see Eq.~\ref{eq:fit} and inset of Fig.~\ref{fig:T1}).
Thirdly, the \cu\ NMR spectra broaden significantly upon lowering the temperature
(see Fig.~\ref{fig:spec}). This broadening is much
stronger than what has been observed in the pure compound \srcuod.\cite{Takigawa1997spectra,Boucher2000} There, a subtle structure of shoulders, broadening and splitting evolves at low temperatures and has been assigned to a field induced local staggered magnetization developing at broken chain ends.\cite{Takigawa1997spectra,Boucher2000} While these chain ends might also be present in the investigated \srcacuod, the associated subtle structure of the spectra cannot be observed. Already at room temperature, the spectrum is broader than the one of \srcuod. This is  due to the impact of disorder induced by the Ca doping, wich is especially demonstrative at the satellites. Disorder mainly affects the quadrupolar interaction between the quadrupolar moment of the nuclei and the disordered electrical field gradient. However, the further NMR line broadening upon lowering the temperature evolves simultaneously for all the resonance lines, suggesting a magnetic origin. In line with the upturn of \slrr\ and the decrease of $\lambda$, this broadening points towards slowly fluctuating magnetic moments.

These results suggest that the spin excitations in the chains of
\srcacuod\ are not simply a subject to a spin gap, but that there
could be regions in the sample which are closer to a magnetic
order. Therefore, the spin gap should rather be regarded as a
pseudo spin gap, in line with our previous DMRG calculations\cite{HammerathPRL2011} as well as with recent inelastic neutron scattering measurements on SrCu$_{0.99}$Ni$_{0.01}$O$_2$.\cite{ Simutis2013}

Let us recall, that the parent compound \srcuod\ exhibits a 3D
magnetic order below $T_N \sim$5\,K.\cite{Keren1993} On the
other hand, \srcuo\ orders only below $T_N \sim 2$\,K. This could
suggest that the low temperature upturn of \slrr\ in \srcacuod\
points towards an onset of magnetic order.

\begin{figure}
\centering
\includegraphics[width=0.9\linewidth]{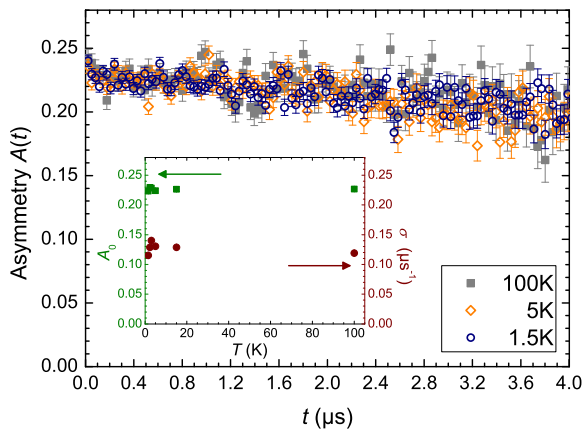}
\caption{\label{fig:asym} ZF $\mu$SR asymmetries of \srcacuod\ measured at 100\,K (grey squares), 5\,K (orange diamonds) and 1.5\,K (blue dots), respectively. The inset shows the temperature dependence of the initial asymmetry $A_0$ (green squares) and the Gaussian decay rate $\sigma$ (brown dots) obtained by fitting the asymmetry time spectra to Eq.~\ref{eq:muSR}.}
\end{figure}  

To search for such a magnetic order at low temperatures, we
performed ZF $\mu$SR measurements on \srcacuod\ down to 1.5\,K.
Due to the extreme sensitivity of the muons to small magnetic
fields, $\mu$SR is ideally suited to study possibly weak and local
magnetism. Some representative ZF $\mu$SR asymmetry spectra are
shown in Fig.~\ref{fig:asym}. For all temperatures (from 100\,K
down to 1.5\,K) we observe only a simple Gaussian decay due to
nuclear dipolar interaction. The decay rate $\sigma$ and the
initial asymmetry $A_0$, obtained by fitting the ZF asymmetry time
dependence $A(t)$ to:
\begin{equation}
\label{eq:muSR}
A(t) =A_0e^{-(\sigma t)^2}
\end{equation} do not change upon cooling down (see inset of Fig.~\ref{fig:asym}).
Additionally, we performed transverse field (TF) $\mu$SR
measurements in 30\,G and could observe the whole initial
asymmetry also at the lowest measured temperature 1.5\,K (not
shown). These results suggest that no magnetic order sets in in
\srcacuod\ down to 1.5\,K in zero external field. This
observation is in line with measurements of the specific heat,
which also did not show any evidence of a magnetic order down to
2\,K.\cite{Beesettyprivate} Nonetheless, the upturn of the \cu\
NMR \slrr, together with the decrease of the stretching exponent
$\lambda$ and the strong NMR line broadening are pointing towards
growing magnetic correlations at least in a strong magnetic
field at low temperatures. These results indicate an approach
towards a magnetic order at lower temperatures than the
experimentally accessed ones, or in high magnetic fields.

The suppression of long range magnetic order by the opening of a gap
is, however, a remarkable observation. To the best of our
knowledge, only the reverse effect has been observed so far,
namely the appearance of magnetic order in a nomally gapped system
induced by impurity doping.\cite{Azuma1997, Grenier2002, Casola2010}

\section{Conclusion}

We have used NMR measurements to investigate the low energy spin
dynamics of the S=1/2 antiferromagnetic Heisenberg spin chain
compound \srcacuod. We found an exponential decrease of the \cu\
NMR \slrr\ upon decreasing temperature, indicating the opening of
a pseudo spin gap of the same order of magnitude as in the double
chain compound \srcacuo. The observation of the spin gap {\it per se} proves that it is solely induced by a local bond disorder causing a small alternation of the intrachain exchange coupling $J$ in Ca doped \srcuo\ and \srcuod, and that it does not depend on the existence of an interchain coupling $J'$. The observation of the very same spin gap value $\Delta$ in both compounds is surprising and might indicate an intrinsic energy scale defined by a collective ordering phenomena in disordered spin chain systems, which should be investigated in more detail in the future. 
At low temperatures, we observe an upturn of \slrr,
accompanied by a growing distribution of \slrr, expressed in a
decreasing stretching exponent $\lambda$ and a simultaneous
broadening of all the \cu\ NMR resonance lines. While we could
prove the absence of any kind of static magnetic order by ZF
$\mu$SR measurements down to 1.5\,K, these NMR features
indicate growing magnetic correlations at low temperatures
in a strong magnetic field. Nevertheless, the evidence for a
suppression of magnetic order and the concomitant appearance of a
spin gap remains unambiguous.

\section*{Acknowledgement}

The assistance by A. Amato and H. Luetkens during the $\mu$SR
measurements at PSI is gratefully acknowledged. The authors thank
R.~Vogel for technical support, S.-L.~Drechsler for discussion and
PSI EU funding for financial support. This work has been supported
by the European Commisssion through the LOTHERM project (Project No.
PITN-GA-2009-238475) and by the Deutsche Forschungsgemeinschaft (DFG) through Grant No. GR3330/4-1 and through the D-A-CH project No. HE3439/12.

\end{document}